\def\be{\begin{equation}}
\def\ee{\end{equation}}
\def\ba{\begin{array}}
\def\ea{\end{array}}

\documentclass[12pt]{article}
\usepackage{CJK}
\usepackage{amssymb}
\usepackage{amsmath}
\usepackage{enumerate}
\usepackage{graphicx}
\usepackage{amsfonts}
\usepackage{url}
\usepackage{bm}
\newtheorem{Proposition}{Proposition}
\newtheorem{Theorem}{Theorem}
\newtheorem{Lemma}{Lemma}
\newcommand\btd{\raise 2pt \hbox{$\hat\bigtriangledown$}\hskip 1.5pt}
\newcommand\bt{\raise 2pt \hbox{$\bigtriangledown$}\hskip 1.5pt}
\begin{document}

\title{\large\bf Geometry of Quantum Computation with Qutrits}
\author{Bin LI$^\dag$$^\ddag$, Zu-Huan YU$^\dag$ and Shao-Ming FEI$^{\ast\dag}$
\\[10pt]
\footnotesize
\small $^\dag$School of Mathematical Sciences, Capital Normal University,\\
\small Beijing 100037, P. R. China\\
\footnotesize \small $^\ddag$School of Mathematics and Statistics, Northeast Normal University,\\
\small Jilin Changchun 130024, P. R. China}
\date{}

\maketitle

\centerline{$^\ast$ Correspondence to feishm@cnu.edu.cn}
\bigskip

\begin{abstract}

Determining the quantum circuit complexity of a unitary operation is an important
problem in quantum computation.
By using the mathematical techniques of Riemannian geometry, we
investigate the efficient quantum circuits in quantum computation
with $n$ qutrits. We show that the optimal quantum circuits are essentially
equivalent to the shortest path between two points in a certain curved geometry of
$SU(3^n)$. As an example, three-qutrit systems are investigated in detail.

\end{abstract}
\bigskip

Due to the quantum parallelism, quantum computers can
solve efficiently problems that are considered
intractable on classical computers \cite{MI}, e.g., algorithm for finding the prime
factors of an integer \cite{PS,P} and quantum searching algorithm \cite{qs}.
A quantum computation can be described as
a sequence of quantum gates, which
determines a unitary evolution $U$ performed
by the computer. An algorithm is
said to be efficient if the number of gates
required grows only polynomially with the size of the problem.
A central problem of quantum computation is to find efficient
quantum circuits to synthesize desired unitary operation $U$
used in such quantum algorithms.

A geometric approach to investigate such quantum circuit complexity for qubit systems has been
developed in \cite{MMMA,M,MMMA2}. It is shown that the
quantum circuit complexity of a unitary operation is closely related
to the problem of finding minimal length paths in a particular curved
geometry. The main idea is to
introduce a Riemannian metric on the space of $n$-qubit unitary
operations, chosen in such a way that the metric distance
$d(I,U)$ between the identity operation and a desired unitary
$U$ is equivalent to the number of quantum gates required to
synthesize $U$ under certain constraints. Hence the distance
$d(I,U)$ is a good measure of the difficulty of synthesizing $U$.

In fact, $d$-dimensional quantum states (qudits) could be more
efficient than qubits in quantum information processing such as
key distribution in the presence of several eavesdroppers.
They offer advantages such as increased
security in a range of quantum information protocols \cite{hsec1,hsec2,hsec3,hsec4,hsec5},
greater channel capacity for quantum communication \cite{hgc}, novel
fundamental tests of quantum mechanics \cite{hnt}, and more
efficient quantum gates \cite{hqg}. In particular, hybrid
qubit-qutrit system has been extensively studied and already
experimentally realized \cite{23e1,23e2}.
The higher dimensional version of qubits provides deeper insights in the
nature of quantum correlations and can be accessed by encoding
qudits in the frequency modes of photon pairs
produced by continuous parametric down-conversion.

In particular, the three-dimensional quantum states, qutrits
are of special significance. For instance, in the
state-independent experimental tests of quantum contextuality,
three ground states of the trapped $^{171}Yb^{+}$ ion are mapped to a
qutrit system and quantum operations are carried out by applying microwaves
resonant to the qutrit transition frequencies \cite{qt}.
The solid-state system, nitrogen-vacancy center in diamond,
can be also served as a qutrit system, in which the electronic
spin can be individually addressed, optically polarized, manipulated and
measured with optical and microwave excitation. Due to its long coherence
time, it is one of the most promising solid state systems as quantum
information processors.

In this paper we study the quantum information processing on qutrit systems.
We generalize the results for qubit-systems \cite{MMMA2} to qutrit ones.
The efficient quantum circuits in quantum computation
with $n$ qutrits are investigated in terms of the geometry of $SU(3^n)$.
Three-qutrit systems are investigated in detail.
Compared with the results for qubit systems \cite{MMMA2}, our results are more fined,
in the sense that by using enough  one\textbf{-} and two\textbf{-}qutrit
gates it is possible to synthesize a unitary operation with
sufficient accuracy. While from \cite{MMMA2}, it is not guaranteed
that the error of the approximation would be arbitrary small.

\medskip
\noindent{\bf Results}

A quantum gate on $n$-qutrit states is a unitary operator
$U\in SU(3^n)$ determined by time-dependent Hamiltonian
$H(t)$ according to the Schr$\ddot{o}$dinger equation,
\begin{equation} \label{eq:1}
\left\{ \begin{aligned}
         \ &\dfrac{dU(t)}{dt}=-iH(t)U(t)\\
         \ &U(0)=I,~~U(T)=U.
 \end{aligned} \right.
\end{equation}

For qutrit case the Hamiltonian $H$ can be expanded in terms of the
Gell-Mann matrices. As the algebra related to the $n$-qutrit space
has rather different properties from the qubits case in which the evolved Pauli matrices
have very nice algebraic relations, we first present some needed results
about the algebra $su(3^n)$.

Let $\lambda_{i}$, $i=1,...,8$, denote the Gell-Mann matrices,
\[ \begin{split}
 \lambda_{1}&=\begin{pmatrix}
0 & 1 & 0\\
1 & 0 & 0\\
0 & 0 & 0
\end{pmatrix},~\lambda_{2}=\begin{pmatrix}
0 & -i& 0\\
i & 0 & 0\\
0 & 0 & 0
\end{pmatrix},~\lambda_{3}=\begin{pmatrix}
1 & 0 & 0\\
0 & -1 & 0\\
0 & 0 & 0
\end{pmatrix},~\lambda_{4}=\begin{pmatrix}
0 & 0 & 1\\
0 & 0 & 0\\
1 & 0 & 0
\end{pmatrix},\\
\lambda_{5}&=\begin{pmatrix}
0 & 0 & -i\\
0 & 0 & 0\\
i & 0 & 0
\end{pmatrix},~\lambda_{6}=\begin{pmatrix}
0 & 0 & 0\\
0 & 0 & 1\\
0 & 1 & 0
\end{pmatrix},~\lambda_{7}=\begin{pmatrix}
0 & 0 & 0\\
0 & 0 &-i\\
0 & i & 0
\end{pmatrix},~\lambda_{8}=\frac{1}{\sqrt{3}}\begin{pmatrix}
1 & 0 & 0\\
0 & 1 & 0\\
0 & 0 &-2
\end{pmatrix}.\\
\end{split} \]
Let
$$
\lambda^{\alpha}_k=I\otimes\cdots\otimes\lambda_{k}\otimes \cdots \otimes I
$$
be an operator acting on the $\alpha$th qutrit with $\lambda_{k}$ and the rest
qutrits with identity $I$.
The basis of $su(3^n)$ is constituted by $\{\Lambda_{s}\}$, $s=1,...,n$, where
$$
\Lambda_{s}=\prod_{i=1}^s(\lambda^{\alpha_i}_{k_i}),
$$
$1\leq\alpha_1<\alpha_2<...<\alpha_s\leq n$, $1\leq k_i\leq 8$.
$\Lambda_{s}$ stands for all operators acting on $s$ qutrits
at sites $\alpha_1,\alpha_2,...,\alpha_s$ with Gell-Mann matrices
$\lambda_{k_1},\lambda_{k_2},...,\lambda_{k_s}$ respectively,
and the rest with identity.
We call an element in $\{\Lambda_{s}\}$ an $s$-body one.
By using the commutation relations among the Gell-Mann matrices, it is
not difficult to prove the following conclusion:

\begin{Lemma}\label{l0}\emph{All $s$-body items ($s\geq3$)
in the basis of $su(3^n)$ can be
generated by the Lie bracket products of $1$-body and
$2$-body items.}
\end{Lemma}

In the following the operator norm of an operator $A$ will be defined by
\be\label{norm}
\|A\|=\displaystyle\max_{\|x\|=1}\|Ax\|,
\ee
which is equivalent to the operator norm given by
$<A,B>=trA^\dag B$. The norm of above Gell-Mann matrices satisfies
$\|\lambda_i\|=1$, $i=1,\cdots,7$, and
$\|\lambda_8\|=\frac{2}{\sqrt{3}}$. If we replace $\lambda_8$ with
$\frac{\sqrt{3}}{2}\lambda_8$, the Gell-Mann matrices are then normal
with respect to the definition of the operator norm, and the basis
of $su(3^n)$, still denoted by $\{\Lambda_s\}$, is normalized.

A general unitary operator $U\in SU(3^n)$
on $n$-qutrit states can be expressed as $U=U_1U_2\cdots U_k$ for some integer $k$.
According to Lemma 1, every $U_i$
acts non-trivially only on one or two vector components of a quantum state vector,
corresponding to a Hamiltonian $H_i$ containing only one and
two-body items in $\{\Lambda_{s}\}$, $s=1,2$.

The time-dependent Hamiltonian $H(t)$ can be expressed as
$$
H=\sum_{\sigma}^{'} h_{\sigma}\sigma+\sum_{\sigma}^{''}
h_{\sigma}\sigma,
$$
where: (1) in the first sum $\sum_{\sigma}^{'}$, $\sigma$ ranges
over all possible one and two-body interactions; (2) in the second
sum $\sum_{\sigma}^{''}$, $\sigma$ ranges over all other more-body
interactions; (3) the $h_{\sigma}$ are real coefficients. We define
the measure of the cost of applying a particular Hamiltonian in
synthesizing a desired unitary operation $U$, similar to the qubit
case,
\begin{equation}\label{fh}
F(H)=\sqrt{\sum_{\sigma}^{'} h_{\sigma}^2+p^2\sum_{\sigma}^{''}
h_{\sigma}^2},
\end{equation}
where $p$ is the penalty paid for applying three- and
more-body items.

Eq. (\ref{fh}) gives rise to a
natural notion of distance in the space $SU(3^n)$
of $n$-qutrit unitary operators with unit determinant.
A curve $[U]$ between the identity
operation $I$ and the desired operation $U$ is a
smooth function,
\begin{equation} \label{eq:2}
\left\{ \begin{aligned}
         \ U&:[0,t_{f}]\rightarrow SU(3^n) \\
         \ U&(0)=I~\mbox{and}~U(t_{f})=U.
        \end{aligned} \right.
\end{equation}
The length of this curve is given by $d([U])\equiv
\int_{0}^{t_{f}}dt F(H(t))$. As d([U]) is invariant with respect to
different parameterizations of $[U]$, one can always
set $F(H(t))=1$ by rescaling $H(t)$, and hence $U$ is
generated at the time $t_f=d([U])$. The distance
$d(I,U)$ between $I$ and $U$ is defined by
\begin{equation} \label{eq:3}
d(I,U)=\min_{\forall[U]}d[U].
\end{equation}

The function $F(H)$ can be
thought of as the norm associated to a right
invariant Riemannian metric whose metric
tensor $g$ has components:
\begin{equation}\label{metric}
g_{\sigma\tau}=\begin{cases}
0 ,& \mbox{if}~\sigma\neq\tau\\
1 ,& \mbox{if}~\sigma=\tau~\mbox{and}~\sigma,~\tau~\mbox{is~ one$\textbf{-}$~or~two$\textbf{-}$body}\\
p^2, & \mbox{if}~ \sigma=\tau~\mbox{and}~ \sigma,~\tau~
\mbox{is~three$\textbf{-}$~or~more$\textbf{-}$body}.
\end{cases}
\end{equation}
These components are written with respect to a basis for local
tangent space corresponding to the coefficients
$h_\sigma$. The distance $d(I,U)$ is equal
to the minimal length solution to the geodesic equation, $\langle dH/dt, J\rangle=i\langle H,[H,J]\rangle$.
Here $\langle\cdot,\cdot\rangle$ is the inner product on the tangent
space $su(3^n)$ defined by the above metric components, and $J$ is
an arbitrary operator in $su(3^n)$.

From Lemma \textbf{1} in the
basis $\{\Lambda_s\}$ of $su(3^n)$, all the $q\textbf{-}$body
items ($q\geq3$) can be generated by Lie bracket products
of $1\textbf{-}$body and $2\textbf{-}$body items. To find the
minimal length solution to the geodesic equation, it is
reasonable to choose such metric (\ref{metric}), because the influence of
there\textbf{-} and more\textbf{-}body items will be ignorable for
sufficiently large $p$. It is the one\textbf{-} and
two\textbf{-}body items that mainly contribute to the minimal geodesic.

We first project the Hamiltonian $H(t)$ onto $H_P(t)$ which contains only one- and two-qutrit
items. By choosing the penalty $p$ large enough we can ensure that the error in
this approximation is small. We then divide the evolution according to $H_P(t)$ into
small time intervals and approximate with a constant mean Hamiltonian over each interval.
We approximate evolution according to the constant mean Hamiltonian
over each interval by a sequence of one- and two-qutrit quantum gates. We show that the total errors
introduced by these approximations can be made arbitrarily smaller than any desired constant.

Let $M$ be a connected manifold and $\mathcal{D}$ a  connection on a principal
$G$-bundle. The Chow's theorem \cite{D} says that
the tangent space $M_q$ at any point $q\in M$ can be divided into
two parts, the horizontal space $H_q M$ and the vertical space $V_q M$, where $M_q=H_q M\oplus V_q M$
and $V_q M\cong \mathfrak{g}$ ($\mathfrak{g}$ denotes the Lie algebra of G). Let $\{X_i^h\}$ be a local
frame of $H_qM$. Then any two points on $M$ can be joined by a horizontal curve if the iterated
Lie brackets $[X_{i_k}^h,[X_{i_{k-1}}^h,\cdots,[X_{i_2}^h,X_{i_1}^h]]\cdots]$ evaluated at $q$ span the tangent space $M_q$.

\begin{Lemma}\label{l1} \emph{Let $p$ be the penalty paid for applying three\textbf{-} and
more\textbf{-}body items. If one chooses $p$ to be sufficiently large,
the distance $d(I,U)$ always has a supremum which is independent of $p$.}
\end{Lemma}

\noindent\textbf{Proof:} As $SU(3^n)$ is a connected and complete manifold,
the tangent space at the identity element $I$ can be looked upon as the Lie algebra $su(3^n)$.
For a given right
invariant Riemannian metric\index{Riemannian metric} (6), there exists a unique geodesic
joining $I$ and some point $U\in SU(3^n)$. With the increase of $p$,
the distance $d(I,U,p)$ of the geodesic
joining $I$ and $U\in SU(3^n)$ increases monotonically.

On the other hand, according to Lemma \textbf{1},  $1\textbf{-}$body and
$2\textbf{-}$body items in the basis $\{\Lambda_s\}$ can span the whole space $su(3^n)$ in terms of the Lie bracket iterations.
Under the metric Eq.(6), from the Chow's theorem  we have that
the horizontal curve joining $I$ and $U\in SU(3^n)$ is unique, since the subspace spanned by $1\textbf{-}$body and
$2\textbf{-}$body items is invariable. Or there exists such a geodesic that its initial tangent vector
lies in the subspace spanned by $1\textbf{-}$body and $2\textbf{-}$body items. Hence the distance $d(I,U,p)$
has a sup $d_0$ which is independent of $p$.
\hfill $\Box$

\begin{Lemma}\label{l2} Let $H_P(t)$ be the projected Hamiltonian containing only
one- and two-body items,
 obtained from a Hamiltonian $H(t)$ generating a unitary operator $U$, and $U_P$
 the corresponding unitary operator generated by $H_P(t)$. Then
 \be\label{L3}
 \|U-U_P\|\leq \dfrac{3^nd([U])}{p},
 \ee
 where $\|\cdot\|$ is the operator norm defined by (\ref{norm}), and $p$ is
 the penalty parameter in (6).
\end{Lemma}

\textbf{Proof:} Let $U$ and $V$ be unitary operators generated by the
time-dependent Hamiltonians $H(t)$ and $J(t)$ respectively,
$$
\dfrac{dU}{dt}=-iHU,~~~~~~\dfrac{dV}{dt}=-iJV.
$$
By integrating above two equations in the interval $[0,T]$, we have
$$
U-V=\int_0^Ti(JV-HU)dt,
$$
where $U(T)=U$, $V(T)=V$ and $U(0)=V(0)=I$ have been taken into account.

Since
$$
\dfrac{dV^\ast U}{dt}=(-iJV)^\ast U+V^\ast(-iHU)
=iV^\ast(J-H)U,
$$
we have
$$
V^\ast U-I=-i\int_0^TV^\ast(H-J)U dt.
$$
Using the triangle inequality and the unitarity of the operator norm $\|\cdot\|,$
we obtain:
\[ \begin{split}
\|U-V\|&=\|V^\ast(U-V)\|=\|V^\ast U-I\|=\|-i\int_0^TV^\ast(H-J)U
dt\|\\
&\leq\int_0^Tdt\|V^\ast(H-J)U\|=\int_0^Tdt\|(H-J)\|.
\end{split} \]

The Euclidean norm of the Hamiltonian $H=\sum_\sigma h_\sigma \sigma$
is given by
$\|H\|_2=\sqrt{\sum_{1}^{N}h_i^2}$.
From the Cauchy-Schwarz inequality, we have
\[\|H\|=\|\sum_\sigma h_\sigma \sigma\|\leq \sum_\sigma |
h_\sigma|\leq3^n\sqrt{h_1^2+h_2^2+\cdots+h_N^2}=3^n\|H\|_2,\]
Moreover, if $H$ contains only three\textbf{-} and more\textbf{-}body items,
we have
\[F(H)=\sqrt{p^2\sum_{\sigma}^{''}h_{\sigma}^2}=p\|H\|_2.\]
Therefore
$$
\ba{rcl}
d([U])&=&\int_0^TdtF(H(t))\\[3mm]
&\geq&\int_0^TdtF(H(t)-H_P(t))=\int_0^T pdt\|H(t)-H_P(t)\|_2\\[3mm]
&\geq&\dfrac{p}{3^n}\int_0^Tdt\|H(t)-H_P(t)\|\geq\dfrac{p}{3^n}\|U-U_P\|,
\ea
$$
which gives rise to (\ref{L3}).
\hfill $\Box$

\noindent\textbf{Remark} From Lemma \ref{l2}, by choosing $p$ sufficiently large, say $p=9^n$, we can
ensure that $\|U-U_P\|\leq d([U])/3^n$. Moreover, since the distance $d(I,U)$ is defined by
$d(I,U)=\min_{\forall[U]}d[U]$, Lemma \textbf{\ref{l2}} also implies that $\|U-U_P\|\leq d(I,U)/3^n.$

\begin{Lemma}\label{l3}
If $U$ is an $n$\textbf{-}qutrit unitary operator generated
by $H(t)$ satisfying
$\|H(t)\|\leq c$ in a time interval $[0,\triangle]$, then
$$
\|U-exp(-i\bar{H}\triangle)\|\leq2(e^{c\triangle}-1-c\triangle)=O(c^2\triangle^2),
$$
where $\bar{H}\equiv\frac{1}{\triangle}\int_0^\triangle dtH(t)$
is the mean Hamiltonian.
\end{Lemma}

\noindent\textbf{Proof:} Recall the Dyson series \cite{J}:
\[U=\sum_{m=0}^\infty (-i)^m \int_{0}^\triangle dt_1\int_{0}^{t_1}
dt_2\cdots\int_{0}^{t_{m-1}} dt_mH(t_1)H(t_2)\cdots H(t_m).\]
We choose $t_i\leq \triangle/(i+1)$ and set
the first term in the above series to be I. Hence the second term is
$(-i)\int_{0}^\triangle H(t_1)dt_1=-i\triangle\bar{H}.$
We have
\[ \begin{split}
&\|e^{(-i\bar{H}\triangle)}-U\|=\|I+(-i\bar{H}\triangle)+
\frac{(-i\bar{H}\triangle)^2}{2}+\cdots+\frac{(-i\bar{H}\triangle)^m}{m!}+\cdots\cdots\\
&~~~~~~~~~~~~~~~~~~~~-\sum_{m=0}^\infty (-i)^m \int_{0}^\triangle dt_1\int_{0}^{t_1}
dt_2\cdots\int_{0}^{t_{m-1}} dt_mH(t_1)H(t_2)\cdots H(t_m)\|\\
&=\|\sum_{m=2}^{\infty}\frac{(-i\bar{H}\triangle)^m}{m!}-\sum_{m=2}^\infty
(-i)^m \int_{0}^\triangle dt_1\int_{0}^{t_1}
dt_2\cdots\int_{0}^{t_{m-1}} dt_mH(t_1)H(t_2)\cdots H(t_m)\|\\
&\leq\sum_{m=2}^\infty(\|\frac{(-i\bar{H}\triangle)^m}{m!}\|+\int_{0}^\triangle
dt_1\int_{0}^{t_1} dt_2\cdots\int_{0}^{t_{m-1}}
dt_m\|H(t_1)H(t_2)\cdots H(t_m)\|)\\
&\leq\sum_{m=2}^\infty(\frac{c^m\triangle^m}{m!}+\frac{c^m\triangle^m}{m!})=2(e^{c\triangle}-1-c\triangle),
\end{split} \]
where we have used the standard norm inequality
$\|XY\|\leq\|X\|\|Y\|,$ the condition $\|H(t)\|\leq c$,
$\int_{0}^\triangle dt_1\int_{0}^{t_1} dt_2\cdots\int_{0}^{t_{m-1}}
dt_m=\triangle^m/m!$
and $t_{m-1}\cdots t_1\triangle\leq \frac{\triangle}{m}\frac{\triangle}{m-1}\cdots\frac{\triangle}{2}\triangle=\frac{\triangle^m}{m!}$.
\hfill $\Box$

\begin{Proposition} If $A$ and $B$ are two unitary operators,
then
$$
\|A^N-B^N\|\leq N\|A-B\|.
$$
\end{Proposition}

\noindent\textbf{Proof:} We begin with $N=2$. It is easy to verify that
$$
\|A^2-B^2\|=\|A^2-AB+AB+B^2\|\leq\|A(A-B)\|+\|(A-B)B\|=2\|A-B\|.
$$
Now suppose that this inequality holds for
$N-1$, $N\geq3$, i.e., $\|A^{N-1}-B^{N-1}\|\leq (N-1)\|A-B\|$.
Then for $N$ we have
\begin{equation*}\begin{split}
\|A^N-B^N\|&=\|A^N-A^{N-1}B+A^{N-1}B-B^N\|\\
&\leq
\|A^{N-1}(A-B)\|+\|(A^{N-1}-B^{N-1})B\|\\
&=\|A-B\|+\|(A^{N-1}-B^{N-1})\parallel\\
&\leq\|A-B\|+(N-1)\|A-B\|=N\|A-B\|.
\end{split}
\end{equation*}
\hfill $\Box$

\begin{Lemma}\label{l4}
\emph{Suppose $H$ is an $n$\textbf{-}qutrit one\textbf{-} and
two\textbf{-}body Hamiltonian whose coefficients satisfy
$|h_\sigma|\leq1.$ Then there is a unitary operator $U_A$ which
satisfies
$$
\|e^{-iH\triangle}-U_A\|\leq c_2 n^2\triangle^3,
$$
and can be synthesized by using at most $c_1n^2/\triangle$
one\textbf{-} and two\textbf{-}tribit gates, where $c_1$ and $c_2$
are constants.}
\end{Lemma}

\noindent\textbf{Proof:} We need a modified version of the Trotter formula
\cite{MI}: let A and B be Hermitian operators, then
$e^{i(A+B)\triangle t} = e^{iA\triangle t}e^{iB\triangle t} +
O(\triangle t^2) .$
We divide the interval $[0,\triangle]$ into $N=1/\triangle$ intervals of
size $\triangle^2$. In every interval, we define a unitary operator
\begin{equation*}
U_{\triangle^2}=e^{-ih_1\sigma_1\triangle^2}e^{-ih_2\sigma_2\triangle^2}\cdots e^{-ih_L\sigma_L\triangle^2}.
\end{equation*}
There are $L=32n^2-24n=O(n^2)$ one\textbf{-} and two\textbf{-}body items in $H$. From the
modified Trotter formula, there exists a constant $c_2$ such that
\begin{equation*}
\begin{split}
\|e^{-iH\triangle^2}-U_{\triangle^2}\|&=\|e^{-i(h_1\sigma_1+h_2\sigma_2+\cdots+h_L\sigma_L)\triangle^2}-
e^{-ih_1\sigma_1\triangle^2}e^{-ih_2\sigma_2\triangle^2}\cdots
e^{-ih_L\sigma_L\triangle^2}\|
\leq c_2n^2\triangle^4.
\end{split}
\end{equation*}

By using Proposition \textbf{1}, we have
$$\|e^{-iH\triangle}-U_{\triangle^2}^N\|\leq N\|e^{-iH\triangle^2}-U_{\triangle^2}\|\leq c_2Nn^2\triangle^4=c_2n^2\triangle^3.$$
It means that one can approximate $e^{-iH\triangle}$ by using at most
$Nc_1n^2=c_1n^2/\triangle$ quantum gates for some constant $c_1$.
\hfill $\Box$

From the above we have our main result:

\begin{Theorem}\label{t2}
\emph{Using $O(n^Kd(I,U)^3)$ $(K\in \mathbb{Z})$ one\textbf{-} and two\textbf{-}qutrit
gates it is possible to synthesize a unitary $U_A$ satisfying
$\|U-U_A\|\leq c$, where $c$ is any constant.}
\end{Theorem}

Theorem \ref{t2} shows that the optimal
way of generating a unitary operator in $SU(3^n)$ is to go along the
minimal geodesic curve connecting I and U.
As an detailed example, we study the three-qutrit systems.
In this case the right invariant Riemannian metric (\ref{metric})
turns out to be a more general one \cite{MM},
$$
\langle H,J \rangle\equiv\frac{tr(H\mathcal {G}(J))}{2\times3^2},
$$
where $\mathcal {G}(J)=s\mathcal {S}(J)+\mathcal {T}(J)+p\mathcal
{Q}(J)$,  $p$ is the penalty parameter and $s$ is the parameter
meaning that one-body Hamiltonians may be applied for free when
it is very small, $\mathcal {S}$ maps the three\textbf{-}qutrit
Hamiltonian to the subspace containing only one\textbf{-}body items,
$\mathcal {T}$ to the subspace containing only two\textbf{-}body
items, and $\mathcal {Q}$ to the subspace containing only
three\textbf{-}body items. According to the properties of the
Gell-Mann matrices, they satisfy $[\mathcal {S},\mathcal
{T}]\subseteq\mathcal {T}$, $[\mathcal {S}, \mathcal
{Q}]\subseteq\mathcal {Q}$, $[\mathcal {T}$, $\mathcal
{Q}]\subseteq\mathcal {T}$.

Set $L=\mathcal {G}(H)$, $S\equiv\mathcal {S}(L)$, $T\equiv\mathcal {T}(L)$ and
$Q\equiv\mathcal {Q}(L)$. From the geodesic equation
$\dot{L}=i[L,\mathcal{F}(L)]$, where $\mathcal{F}=\mathcal {G}^{-1}$, we have
\begin{equation} \label{eq:5}
\left\{ \begin{aligned}
\dot{S}&= 0, \\
\dot{T}&=i[(1-\frac{1}{s})S+(1-\frac{1}{p})Q,T],\\
\dot{Q}&=i(\frac{1}{p}-\frac{1}{s})[S,Q],
\end{aligned} \right.
\end{equation}
which gives rise to the solution
\begin{equation} \label{eq:6}
\left\{ \begin{aligned}
\ S(t)&=S_0 \\
\ T(t)&=e^{it(p^{-1}-s^{-1})S_0}e^{it(1-p^{-1})(S_0+Q_0)}T_0e^{-it(1-p^{-1})(S_0+Q_0)}e^{-it(p^{-1}-s^{-1})S_0}\\
\ Q(t)&=e^{it(p^{-1}-s^{-1})S_0}Q_0e^{-it(p^{-1}-s^{-1})S_0}.
\end{aligned} \right.,
\end{equation}
where $S(0)=S_0$, $T(0)=T_0$ and $Q(0)=Q_0$.

 The corresponding Hamiltonian $H=\mathcal {G}^{-1}(L)$ has the
form: $H(t)=\dfrac{1}{s}S(t)+T(t)+\dfrac{1}{p}Q(t).$ According to
the assumption $\langle H(t),H(t)\rangle=1$ for all time $t$, we
have $\dfrac{tr(S^2)}{2\times3^2}\leq
s,~\dfrac{tr(T^2)}{2\times3^2}\leq 1$, and
$\dfrac{tr(Q^2)}{2\times3^2}\leq p$. The term $\dfrac{1}{p}Q(t)$ in
$H(t)$ is of order $p^{-1/2}$, and hence can be neglected in the
large $p$ limit, with an error of order $tp^{-1/2}.$ Also the term
containing $p^{-1}$ in the exponentials of $T$ can be neglected with
an error at most of order $t^2(s^{1/2}p^{-1}+p^{-1/2})$. Therefore
one can define an approximate Hamiltonian
$$
\tilde{H}(t)=\frac{1}{s}S_0+e^{-its^{-1}S_0}e^{it(S_0+Q_0)}T_0e^{-it(S_0+Q_0)}e^{its^{-1}S_0}.
$$
The corresponding solution $\tilde{U}(t)$ of the
Schr$\ddot{o}$dinger equation satisfies
$$
\|U(t)-\tilde{U}(t)\|\leq O(tp^{-1/2}+t^2(s^{1/2}p^{-1}+p^{-1/2})).
$$

Denote
$\tilde{V}=e^{-it(S_0+Q_0)}e^{its^{-1}S_0}\tilde{U}$.
Then
$\dot{\tilde{V}}=-i(S_0+T_0+Q_0)\tilde{V}$ and
$\tilde{V}=e^{-it(S_0+T_0+Q_0)}$. Thus we have
$$
\tilde{U}(t)=e^{-its^{-1}S_0}e^{it(S_0+Q_0)}e^{-it(S_0+T_0+Q_0)}.
$$

Generally one can expect that $S_0+Q_0$ is much lager than $T_0$,
and $S_0+Q_0$ is non-degenerate. $\tilde{U}$ can be simplified
at the first-order perturbation,
$$
\tilde{U}(t)=e^{-its^{-1}S_0}e^{-it\mathcal {R}_{(S_0+Q_0)}(T_0)},
$$
where $\mathcal {R}_{(S_0+Q_0)}(T_0)$ denotes the diagonal matrix
by removing all the off-diagonal entries from $T_0$ in the eigenbasis of $S_0+Q_0$.
Therefore we see that it is possible to synthesize a unitary $\tilde{U}$
satisfying $\|U(t)-\tilde{U}(t)\|\leq c$, where $c$ is any constant, say $c=1/10$.

\medskip
\noindent{\bf Discussions}

We have investigated the efficient quantum circuits in quantum computation
with $n$ qutrits in terms of Riemannian geometry. We have shown that the optimal quantum circuits are essentially
equivalent to the shortest path between two points in a certain curved geometry of
$SU(3^n)$, similar to the qubit case where the geodesic in $SU(2^n)$ is involved \cite{MMMA2}.
As an example, three-qutrit systems have been investigated in detail.
Some algebraic derivations involved for qutrit systems are rather different from the ones in qubit systems.
In particular, we used (\ref{norm}) as the norm of operators.
The operator norm of $M$ used in \cite{MMMA2} is defined by
$\|M\|_1=\max_{\langle\psi|\psi\rangle=1} \{|\langle\psi| M |\psi\rangle|\}$, which is not
unitary invariant in the sense that $\|M\|_1=\|U\,M\|_1=\|M\,U\|_1$
is not always true for any unitary operator $U$. For instance, consider
$M=\begin{pmatrix}
         0 & 1 \\
         0 & 0\\
       \end{pmatrix}$ and
$U=\frac{1}{\sqrt{2}}\begin{pmatrix}
         1 &-i \\
         i & -1\\
       \end{pmatrix}$.
One has $\|M\|_1={1}/{2}$. However, $\|M\,U\|_1={1}/{2}+{\sqrt{2}}/{4}$.
Generally, from Cauchy-Schwarz inequality one has $\|M \|_1\leq \|M \|$.
If $M^\dag M=I$ or $M^\dag=M$, then $\|M\|_1=\|M\|$.

Moreover, the final results we obtained are finer than the ones
in \cite{MMMA2}. Our result shows that if $k$ in formula $\dfrac{1}{\triangle}=n^kd(I,U)$ is taken to be sufficiently large,
$\|U-U_A\|$ can be sufficiently small.
However, the approximation error estimation in \cite{MMMA2} reads
$$
||U-U_A||\leq\frac{d(I,U)}{2^n}+2\frac{d(I,U)}{\Delta}(e^{(3/\sqrt{2})n\Delta}-(1+\frac{3}{\sqrt{2}}n\Delta))+c_2d(I,U)n^4\Delta^2.
$$
First, since $d(I,U)$ is dependent of the penalty parameter $p$, there should exist a $p$-independent bound
to guarantee that ${2^n d(I,U)}/{p}$ is small for sufficiently large $p$.
Second, if one chooses $\Delta$ as scale $1/n^2d(I,U)$, the sum of the last two
terms of the right hand side is $9/2+c_2/d(I,U)+ O$. Therefore the scale should be smaller, for example, $1/n^kd(I,U)$ and $k>3$.
As $\triangle$ takes the scale of ${1}/{n^2d(I,U)}$ in \cite{MMMA2}, it can not guarantee
that the error in the approximation could be arbitrary small.

Due to the special properties of the Pauli matrices involved in qubit systems,
many derivations for qubit systems are different from the ones for qutrit systems.
Nevertheless, the derivations for qutrit systems in this paper can be generalized
to general high dimensional qudit systems.

\medskip
\noindent{\bf Methods}

In deriving Theorem \ref{t2}, we use Lemmas 2-5. Let $H(t)$ be the time-dependent normalized
Hamiltonian generating the minimal geodesic of length $d(I,U)$.
Let $H_P(t)$ be the projected Hamiltonian which
contains only the one- and two-body items in $H(t)$ and generates $U_P$.
According to Lemma \ref{l2}, they satisfy
\begin{equation}\label{8}
\|U-U_P\|\leq \frac{3^nd(I,U)}{p}.
\end{equation}

Divide the time interval $[0,d(I,U)]$ into $N$
parts with each of length $\triangle=d(I,U)/N$. Let $U_P^j$ be
the unitary operator generated by $H_P(t)$ in the $j$th time
interval, and $U_M^j$ be the unitary operator generated by the mean
Hamiltonian
$\bar{H}=\frac{1}{\triangle}\int_{(j-1)\triangle}^{j\triangle}
dtH_P(t).$ Then using Lemma \ref{l3} and inequality $\|H_P^j(t)\|\leq4\sqrt{2}n$ we have
\begin{equation}\label{9}
\|U_P^j-U_M^j\|\leq
2(e^{4\sqrt{2}n\triangle}-(1+4\sqrt{2}n\triangle)).
\end{equation}
As $F(H)$ is scaled to be one,
$F(H)=\sqrt{\sum_{\sigma\prime}h_\sigma^2+p^2\sum_{\sigma\prime\prime}h_\sigma^2}=1$,
one has $F(H_{P}(t))=\sqrt{\sum_{\sigma\prime}h_\sigma^2}\leq 1$. Hence
$$
\|H_P(t)\|=\|\sum_{\sigma}^{'}  h_\sigma\sigma\|\leq\sum_{\sigma}^{'}|h_\sigma|
\leq L\sqrt{h_1^2+h_2^2+\cdots\cdots+h_L^2}\leq4\sqrt{2}n,
$$
where $L=32n^2-24n$ is the number of one\textbf{-} and two\textbf{-}body
items in $H(t)$, i.e. the number of the terms in $H_P(t)$).

Applying Lemma \ref{l4} to $\bar{H}^j$ on every time interval, we have that there
exists a unitary $U_A^j$ which can be synthesized by using at most
$c_1n^2/\triangle$ one\textbf{-} and two\textbf{-}qutrit gates, and
satisfies
\begin{equation}\label{10}
\|U_M^j-U_A^j\|=\|e^{-i\bar{H}^j\triangle}-U_A^j\|\leq
c_2n^2\triangle^3.
\end{equation}

$U_P$ and $U_A$ can be generated in terms of $U_P^j$ and $U_A^j$, respectively.
We show how to generate $U_P$ by use of $H_P^j$ below.
First, $U_P^1$ can be generated by $H_P^1$:
\begin{equation*} \label{eq:2p}
\dfrac{dU_P^1}{dt}=-iH_P^1(t)U_P^1(t),~~~~U_P^1(0)=I\,
\end{equation*}
with $U_P^1(\triangle)=U_P^1$. The unitary
operator $U_P^2$ generated by $H_P^2$ satisfies
\begin{equation*}
\label{eq:3}
\dfrac{dU_P^2(t)}{dt}=-iH_P^2(t)U_P^2(t),~~~~U_P^2(\triangle)=U_P^1,
\end{equation*}
which can be transformed into
\begin{equation*}
\label{eq:4}
\dfrac{dU_P^2U_P^1(t)}{dt}=-iH_P^2(t)U_P^2(t)U_P^1(t),~~~~U_P^2U_P^1(0)=U_P^1,
\end{equation*}
with $U_P^2(2\triangle)U_P^1=U_P^2U_P^1$, where $U_P^1$ is constant in $[\triangle,2\triangle]$.
At last we have
$U_P=U_P^NU_P^{N-1}\cdots U_P^1$ generated by the Hamiltonians
$H_P^1(t)$, $H_P^2(t),\cdots,H_P^{N}(t)$.
$U_A$ can be generated similarly.

Therefore
\begin{equation*}\begin{split}
&\|U_P-U_A\|\\
&=\|U_P^NU_P^{N-1}\cdots U_P^1-U_A^NU_A^{N-1}\cdots
U_A^1\|\\
&=\|U_P^NU_P^{N-1}\cdots U_P^1-U_P^NU_P^{N-1}\cdots
U_A^1+U_P^NU_P^{N-1}\cdots U_A^1-U_A^NU_A^{N-1}\cdots U_A^1\|\\
&\leq\|U_P^NU_P^{N-1}\cdots
U_P^2(U_P^1-U_A^1)\|+\|(U_P^NU_P^{N-1}\cdots
U_P^2-U_A^NU_A^{N-1}\cdots U_A^2)U_A^1\|\\
&=\|U_P^1-U_A^1\|+\|U_P^NU_P^{N-1}\cdots U_P^2-U_A^NU_A^{N-1}\cdots U_A^2\|=\cdots\\
&\leq \|U_P^1-U_A^1\|+\|U_P^2-U_A^2\|+\cdots+\|U_P^N-U_A^N\|
=\sum_{j=1}^N \|U_P^j-U_A^j\|.
\end{split}\end{equation*}

From (\ref{8}), (\ref{9}) and (\ref{10}) we obtain:
\begin{equation}\begin{split}
\|U-U_A\|&\leq \|U-U_P\|+\|U_P-U_A\|\\
&\leq \frac{3^nd(I,U)}{p}+\sum_{j=1}^N \|U_P^j-U_A^j\|\\
&\leq\frac{3^nd(I,U)}{p}+\sum_{j=1}^N( \|U_P^j-U_M^j\|+\|U_M^j-U_A^j\|)\\
&\leq\frac{3^nd(I,U)}{p}+2N(e^{4\sqrt{2}n\triangle}-(1+4\sqrt{2}n\triangle))+c_2Nn^2\triangle^3\\
&=\frac{3^nd(I,U)}{p}+2\frac{d(I,U)}{\triangle}(e^{4\sqrt{2}n\triangle}-(1+4\sqrt{2}n\triangle))+c_2d(I,U)n^2\triangle^2\\
&=\frac{3^nd(I,U)}{p}+2c_0d(I,U)n^2\triangle+c_2d(I,U)n^2\triangle^2,
\end{split}\end{equation}
where
$e^{4\sqrt{2}n\triangle}-(1+4\sqrt{2}n\triangle)=O(n^2\triangle^2)$
and $c_0$ is a constant.

As mentioned in Lemma \ref{l1}, the distance $d(I,U)$ has a sup
$d_0$ for sufficiently large $p$. For example, we
choose a suitable penalty $p$ so that $d(I,U,p)$ satisfies
$\dfrac{8d_0}{9}\leq d(I,U,p)\leq d_0$. If we choose $\triangle$ to be
sufficiently small, e.g.
$\dfrac{1}{\triangle}=n^kd(I,U)$ with $k$ sufficiently large,
$\|U-U_A\|$ will be sufficiently small,
\begin{equation}\begin{split}
\|U-U_A\|&\leq \frac{3^nd(I,U)}{p}+2c_0n^{-(k-2)}+\frac{c_2n^{-(2k-2)}}{d(I,U)}\\
&\leq \frac{3^nd_0}{p}+2c_0n^{-(k-2)}+\frac{9c_2}{8d_0}n^{-(2k-2)}.
\end{split}
\end{equation}

As we need $c_1n^2/\triangle$ one\textbf{-} and
two\textbf{-}body gates to synthesize every $U_A^j$, we
ultimately need $\dfrac{c_1n^2}{\triangle}
N=\dfrac{c_1n^2d(I,U)}{\triangle^2}=c_1d(I,U)^3n^{2k+2}$
one\textbf{-} and two\textbf{-}body gates.

\bigskip
\noindent{\sf Acknowledgement}
The work is supported by NSFC under number 11275131.

\newpage
\bigskip
\noindent{\sf Acknowledgements}

\noindent The work is supported by NSFC under number 11275131.

\bigskip
\noindent{\sf Author contributions}

\noindent B.L and Z.H. and S.M. wrote the main manuscript text. All authors reviewed the manuscript.

\bigskip
\noindent{\sf Additional Information}

\noindent Competing Financial Interests: The authors declare no competing financial interests.

\end{document}